\documentclass[lettersize,journal]{IEEEtran}
\usepackage{amsmath,amsfonts}
\usepackage{algorithmic}
\usepackage{algorithm}
\usepackage{array}
\usepackage[caption=false,font=normalsize,labelfont=sf,textfont=sf]{subfig}
\usepackage{textcomp}
\usepackage{stfloats}
\usepackage{url}
\usepackage{verbatim}
\usepackage{graphicx}
\usepackage{cite}
\begin{document}

\title{Photonic Filter for THz via optical injection into a\\ feedback-controlled on-chip multi-wavelength laser}

\author{Garzan Arda Akin, Luis Gonzalez-Guerrero, Pablo Marin-Palomo
        % <-this % stops a space
\thanks{The work of Pablo Marin-Palomo is supported by the European Research Council (ERC, Starting Grant COLOR’UP 948129) and the METHUSALEM program of the Flemish Government (Vlaamse Overheid). Pablo Marin-Palomo is a postdoctoral fellow of the Research Foundation Flanders (FWO)  under Grant 1275924N. The work of Luis Gonzalez-Guerrero was supported by the Horizon Europe research and innovation program under Grant No. 101096949, TERA6G project. (Corresponding author: Pablo Marin-Palomo.)}% <-this % stops a space
\thanks{Pablo Marin-Palomo and Garzan Arda Akın are with Brussels Photonics Team (B-PHOT), Vrije Universiteit Brussel (VUB), 1050 Brussel, Belgium (e-mail: Pablo.Marin-Palomo@vub.be ; Garza.Akin@vub.be).}
\thanks{Luis Gonzalez-Guerrero is with Electronic Technology Department, Carlos III University of Madrid, Leganés, Spain (e-mail: lgguerre@ing.uc3m.es).}}

\markboth{Manuscript submitted to JLT for review. September 2025}%
{Shell \MakeLowercase{\textit{et al.}}: A Sample Article Using IEEEtran.cls for IEEE Journals}

\IEEEpubid{0000--0000/00\$00.00~\copyright~2025 IEEE}

\maketitle

\begin{abstract}
The growing demand for ultra-fast wireless data transmission is accelerating the exploration of THz frequencies for next-generation wireless communication systems. THz photonics provides a platform for generating, processing, and detecting THz signals by upconversion to the optical domain and leveraging the large available bandwidth of photonic devices at optical frequencies. However, efficient, tunable, and agile photonic filters remain a significant technical challenge. Here, we present a monolithically-integrated, feedback-controlled multi-wavelength laser (MWL) enabling photonic filtering through regenerative amplification induced by optical signal injection. The feedback cavity enables agile control of the filter response by tuning the feedback phase. We experimentally demonstrate selective filtering of an upconverted THz signal comprising two channels separated by 34 GHz and offset by 1.3~THz from the carrier. Using a broadband signal, our photonic filter achieves up to 20dB suppression ratio, 3-dB bandwidths as narrow as 160 MHz, and optical gain of up to 15~dB for low-power optical signals commonly obtained after THz-to-optical conversion, making it highly attractive for THz communication links. 
\end{abstract}


\section{Introduction}
\IEEEPARstart{T}{he} 
growing demand for higher data transmission rates and the increasing use of THz signals (100 GHz to 10 THz) in spectroscopy, imaging, and sensing applications motivate research in mm-wave and THz technologies \cite{xie_review_2021,huang_terahertz_2023}. These systems require the ability to generate, measure, process, and manipulate ultra-high frequency signals. However, the progress towards THz systems is hindered by the limited access to devices operating at THz, referred to as the THz gap. THz filtering, in particular, is key for sensing, imaging, and communication applications, e.g., for channel selection in high-capacity THz wireless communication systems \cite{ma_frequency-division_2017}.

In recent years, several THz filtering approaches have been proposed \cite{ma_frequency-division_2017,huang_actively_2020,zhang_electrically_2022,dehghanian_demonstration_2023,dong_versatile_2022,fu_graphene-based_2024}. In \cite{ma_frequency-division_2017}, a leaky-waveguide is used to couple the THz signal into free space, where the signal is spatially filtered. In \cite{huang_actively_2020}, a THz filter at 1.8 THz center frequency based on a MEMS-actuated metamaterial was introduced. However, these systems are slow, require high voltages, and suffer from limited tunability. In \cite{zhang_electrically_2022}, liquid crystals in a waveguide cavity showed broad tunability of the center frequency at around 900 GHz. However, the response is slow, its 3-dB bandwidth is rather large, and it is especially sensitive to temperature. THz Bragg gratings have allowed filtering at center frequencies of up to 800 GHz using apodized gratings \cite{dehghanian_demonstration_2023} and 3-dB bandwidths down to 1~GHz by periodically etching a two-wire waveguide \cite{dong_versatile_2022}. However, THz Bragg filters’ tunability range is limited, and there are few results of narrow-band filters with center frequency above 500 GHz. Graphene-based resonators \cite{fu_graphene-based_2024} show transmission peaks centered at several THz, enabling fast tuning and compact integration, but face challenges due to large insertion losses, low efficiency, and fabrication complexity. Overall, these THz filter techniques are, in general, harder to integrate with sources and detectors, suffer from large losses, and are inherently limited in tunability and spectral selectivity, i.e., feature a large full width at half maximum (FWHM).

Microwave photonic (MWP) filters for THz signal processing have recently emerged as a promising alternative to electronic THz filters. In MWP, the RF signal, which lies, in this case, in the THz range, is first upconverted into the optical domain using a modulator, processed with photonic components, and then converted back into the RF domain using a photodiode (PD). Compared with conventional RF filters, MWP filters offer large center-frequency tuning ranges and high levels of reconfigurability. In addition, the development of integrated photonic platforms enables their implementation in compact and lightweight photonic chips \cite{liu_integrated_2020}. 

MWP filters are generally classified into two groups – incoherent and coherent \cite{yao_photonics_2015}- depending on the optical source used for upconversion. Incoherent MWP filters are based on tapped delay-line architectures, where the input RF signal is sampled, delayed, and weighted before being summed \cite{capmany_discrete-time_2005}. These filters exhibit a periodic frequency response. Thus, in order to suppress spurious signals at lower frequencies, the spacing between spectral replicas should be in the THz. This can be achieved using arrays of lasers or with optical frequency comb generators producing decorrelated optical lines \cite{capmany_discrete-time_2005,xu_advanced_2019}. However, both approaches significantly increase the complexity of the MWP filter. Coherent MWP filters use low-linewidth lasers and optical filters, mapping the response of the optical filter directly to the RF domain \cite{liu_integrated_2020}. 
\IEEEpubidadjcol

Scaling the center frequency of these filters into the THz domain is relatively straightforward, as it can be achieved by tuning the laser wavelength until the modulation sideband to be filtered coincides with the center frequency of the optical filter. The optical filters employed in coherent MWP architectures are typically based on coupled resonator optical waveguide (CROW) structures implemented on low-loss integration platforms such as silicon nitride, enabling the realization of very high-Q responses \cite{morichetti_first_2012}. However, CROW filters require cascading many resonators to achieve MHz-range bandwidths \cite{taddei_high-selectivity_2019}, which increases chip size and tuning complexity. Moreover, since tuning typically relies on thermal heaters, they suffer from thermal crosstalk, slow reconfiguration speeds, and high power consumption, making large-scale implementations inefficient. Both groups can handle THz signals, owing to recent advances in optical modulators and PDs with optoelectronic bandwidths extending into the THz range \cite{ummethala_thz--optical_2019,ohara_2-mw-output_2023}. Nonetheless, several implementation challenges remain when scaling MWP filters to operate effectively in the THz regime.

In this work, we introduce a narrowband, tunable photonic filter approach for MPW THz filtering of signals beyond 1 THz. Our approach obtains extinction ratio levels of up to 20~dB and, through regenerative amplification, our filter provides more than 15~dB of optical gain. The operating principle of our photonic filter relies on optical injection of the signal into a custom-designed integrated multi-wavelength laser (MWL) with a phase-tunable monolithically integrated feedback cavity \cite{virte_integrated_2023,abdollahi_agile_2024}. By tuning the phase of the feedback into the MWL, we achieve agile selection and regenerative amplification of the desired spectral channel. We demonstrate selective filtering of a two-channel optical signal, offset by 1.3~THz from the carrier and with a channel separation of 34~GHz, achieving high suppression ratios, narrow bandwidth, and positive optical gain, well-suited for next-generation THz communication systems. In addition, we analyze the stability of our approach with respect to the signal power and injection detuning, and we demonstrate that the filtering approach can be extended to other THz frequencies.

\section{Photonic Filter Concept and MWL}

\begin{figure}[!t]
\centering
\includegraphics[width=3.5in]{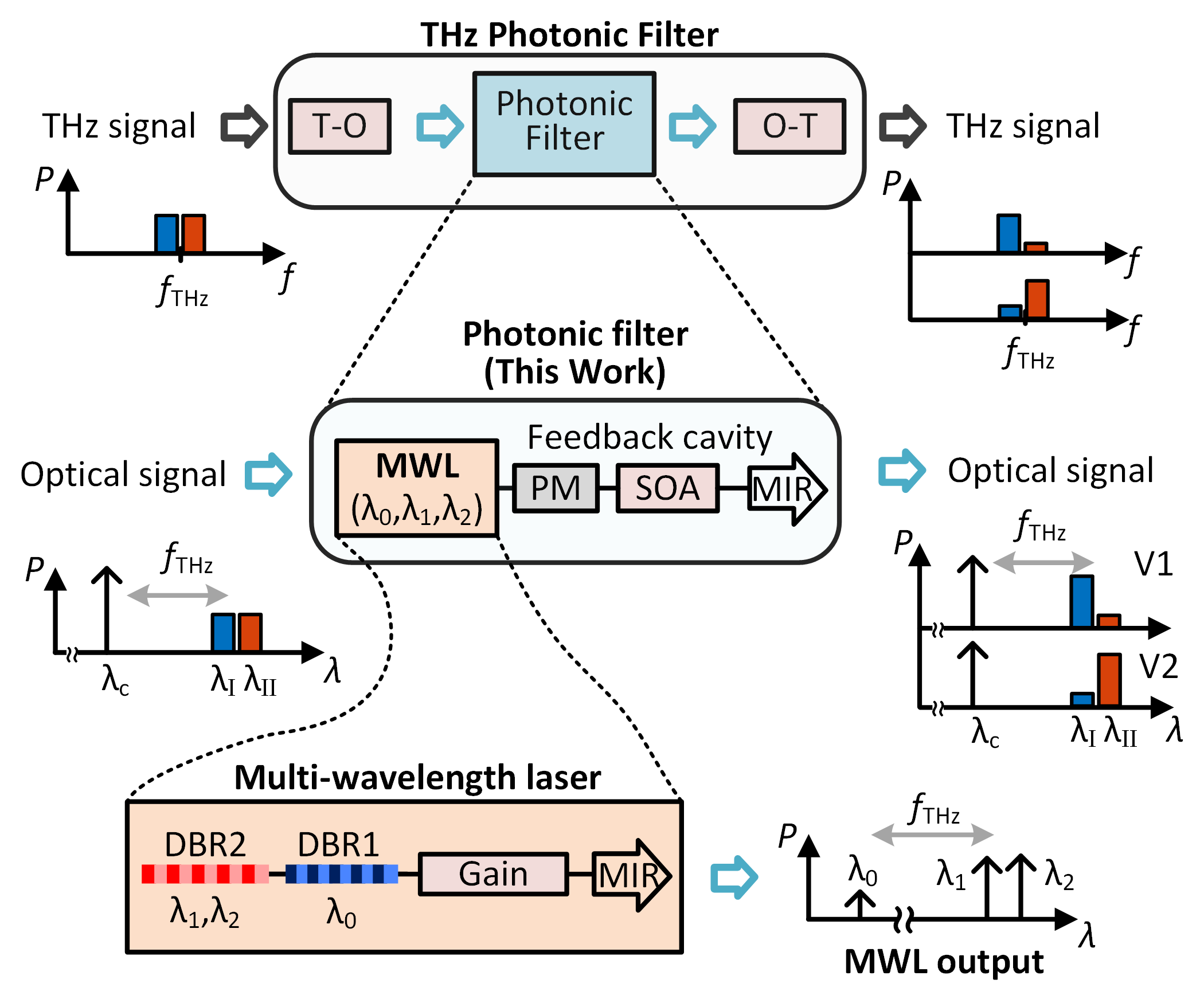}
\caption{Concept of the THz photonic filter using a feedback-controlled multi-wavelength laser (MWL) as filter mechanism. The cavity of the MWL consists of two detuned distributed Bragg reflectors (DBR) on one side and a broadband multimode interference reflector (MIR) on the other side. The Bragg wavelengths of the DBRs are separated by fTHz. The output of the MWL consists of three modes. The dominant modes, $\lambda_1$ and $\lambda_2$, arise from the cavity defined by DBR2, while the suppressed mode at $\lambda_0$ is associated with the cavity defined by DBR1. After THz-to-optical (T-O) conversion with a carrier at $\lambda_\text{c}$, the optical signal is injected into the MWL with $\lambda_\text{c}$ close to $\lambda_0$ to lock the MWL. In these settings, the center wavelength $\lambda_\text{I}$ and $\lambda_{\text{II}}$ of the two THz channels coincide with $\lambda_1$ and $\lambda_2$, respectively, leading to regenerative amplification of the channels. The feedback phase, controlled by applying voltage to a phase modulator (PM), enables adjusting the power of each channel, while the semiconductor optical amplifier (SOA) allows adjusting the feedback strength. The optical signal from the MWL is sent to an optical-to-THz (O-T) converter to obtain the filtered THz signal.}
\label{fig_1}
\end{figure}

Fig. 1 illustrates the operating principle of the proposed photonic filter. The core of our photonic filter is an integrated multi-wavelength laser (MWL). The MWL is defined by a dual-cavity structure, with two distributed Bragg reflectors (DBR) placed in series on one side and a broadband multimode interference reflector (MIR) on the other side \cite{pawlus_control_2023}. The pitch of the DBRs is set so that the Bragg wavelengths are detuned from each other by $f_{\text{THz}}$ (100 GHz to 10 THz). The MWL is operated so that its output consists of three modes. The dominant modes, at $\lambda_1$ and $\lambda_2$, arise from the cavity defined by DBR2, while the suppressed mode at $\lambda_0$ is associated with the cavity defined by DBR1. 
The THz signal, e.g., comprising two channels centered at $f_{\text{THz}}$, to be processed, is first upconverted to $\lambda_\text{I}$ and $\lambda_{\text{II}}$ via THz-to-optical (T-O) conversion. This is achieved by encoding the THz signal onto an optical carrier at $\lambda_\text{c}$ using, e.g., an antenna-coupled high-speed modulator \cite{ummethala_thz--optical_2019}. The resulting optical signal is then injected into our MWL. Note that there may be a non-zero detuning $\delta = \lambda_\text{c} – \lambda_0$ and therefore the wavelengths $\lambda_1$ and $\lambda_2$, emitted by the free-running MWL, may differ from $\lambda_\text{I}$ and $\lambda_{\text{II}}$. When $\lambda_\text{c}$ is chosen to be within the locking range of the MWL, the MWL is locked to the carrier and will emit at $\lambda_\text{c}$ \cite{wieczorek_dynamical_2005}. In this setting, $\lambda_\text{I}$ and $\lambda_{\text{II}}$ will coincide with $\lambda_1$ and $\lambda_2$, respectively, leading to regenerative amplification of the channels \cite{simpson_enhanced_1997}. Here, we use indices 0, 1, and 2 to refer to the MWL modes, while the indices I and II refer to the center wavelengths of each of the upconverted THz channels. Naturally, this approach can be scaled to a higher number of THz channels by proper design of the MWL.

An external cavity monolithically integrated with the MWL and composed of a phase modulator (PM), a semiconductor optical amplifier (SOA), and an MIR is used to apply feedback to the MWL, see Fig. 1. The feedback phase, controlled by applying voltage to the PM, enables adjusting the power of each channel based on destructive or constructive interferences between the intracavity field and the feedback fields \cite{pawlus_control_2023,ladouce_-chip_2024}. Such a feedback control mechanism has been recently shown to be able to switch the emission of the MWL at nanosecond time scales \cite{ladouce_-chip_2024}. Through feedback and with appropriate feedback control, the selected channel is amplified by the MWL with respect to the remaining channels, achieving a large (>10 dB) suppression ratio. In addition, the optical carrier remains at the output of the MWL, and it might be amplified through injection locking, which allows it to be used as the local oscillator during the optical-to-THz (O-T) conversion. The fact that the same carrier is used for both T-O and O-T conversion ensures that the coherence of the signal is maintained during the filtering, i.e., the noise from the semiconductor MWL will not be transferred to the THz signal.

In this work, unless otherwise stated, we use an MWL where the Bragg wavelengths of the DBRs are detuned from each other by approximately 10~nm (1.3~THz), see details in \cite{abdollahi_agile_2024}. We drive the laser at 40~mA, which is approximately twice the lasing threshold, and operate it at a temperature of 22~°C. In this configuration, the MWL emits at $\lambda_0 = 1537.46$~nm, $\lambda_1 = 1547.95$~nm, $\lambda_2 = 1548.23$~nm.

\section{Photonic Filter for 1.3 THz Data Signals}

The experimental setup to perform the proof-of-concept demonstration is depicted in Fig.~2(a). Due to the lack of suitable high-speed modulators, we bypass the T-O conversion and directly emulate the T-O converted signal by using two external-cavity lasers (ECL), one set at $\lambda_\text{c}$ = 1537.475~nm and the other emitting with a 1.3~THz offset with respect to $\lambda_\text{c}$. The former emulates the optical carrier involved in the T-O conversion, while the latter is sent through a Mach-Zehnder modulator (MZM) to generate the two data channels centered at $\lambda_\text{I}$ = 1547.953~nm and $\lambda_{\text{II}}$ = 1548.227~nm, separated by approximately 34~GHz. 

We consider, for simplicity, on-off-keying (OOK) signals generated using an arbitrary waveform generator (AWG, Keysight M8194A). We make use of a pseudo-random bit sequence of length $2^{11}-1$ at a symbol rate of 200~MBd and use a root-raised cosine pulse shaping with a 10\% roll-off. To generate the data channels at $\lambda_\text{I}$ and $\lambda_{\text{II}}$, we digitally upconvert the OOK signal to a frequency of 17~GHz so that a double sideband signal is generated, where the two sidebands are the two corresponding data channels at $\lambda_\text{I}$ and $\lambda_{\text{II}}$. In this configuration, the two channels carry the same information. 

The optical power of the signal is adjusted using a variable optical attenuator (VOA), and a polarization controller (PC) is used to match the polarization of both the signal and the optical carrier before combining them, see Fig.~2(b) for the optical spectrum of the injection signal. In this setting, the on-chip carrier power Pc is 3~dBm while the on-chip signal power $P_{\text{sig}} = -46$~dBm. Such low signal power levels are typical after T-O conversion, cf. \cite{ummethala_thz--optical_2019}. 
The polarization of the combined carrier and signal is adjusted with a PC to align with the polarization of the MWL in the photonic integrated circuit (PIC). The signal is coupled into the PIC using a lensed fiber (LF) with approximately 3~dB of coupling losses and is optically injected with a detuning of approximately 20~pm between $\lambda_\text{c}$ and $\lambda_0$, so that the MWL is locked to the optical carrier. In this setting, the channels’ wavelengths coincide with those of the MWL, enabling regenerative amplification.

\begin{figure}[!t]
\centering
\includegraphics[width=3.5in]{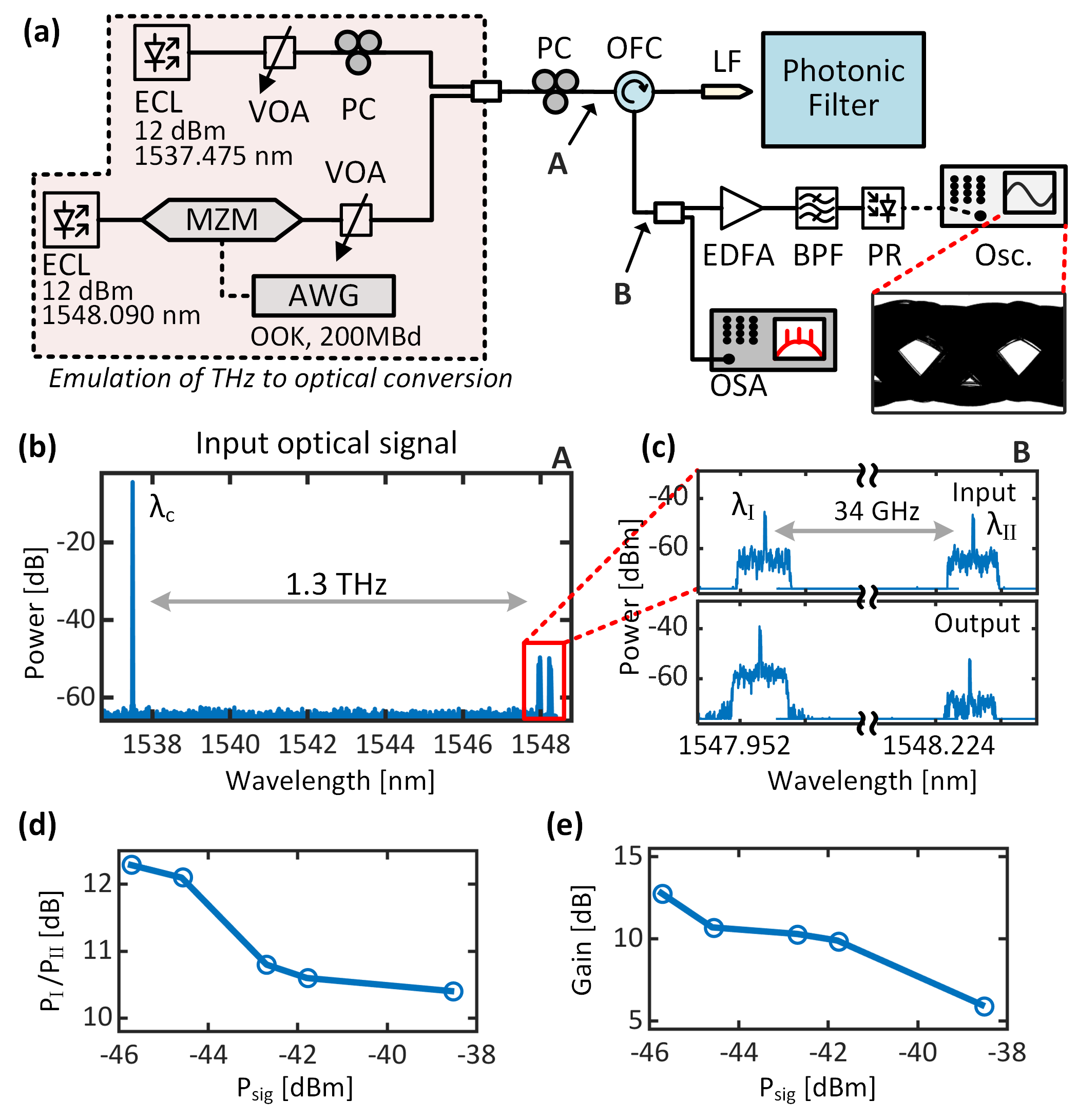}
\caption{Photonic filtering with an MWL. (a) Setup for the generation and filtering of the optical signal emulating THz-to-Optical conversion. Inset: exemplary eye diagram corresponding to a BER below $10^{-10}$ (b) Optical spectrum of the THz-to-optical converted signal measured at A. (c) Top: Zoomed-in optical spectrum of the injection signal. Bottom: Optical spectrum of the signal filtered by our MWL, measured at B. (d) Suppression ratio and (e) Optical power gain of the selected channel at $\lambda_1$. ECL: External-cavity laser. VOA: Variable optical attenuator. MZM: Mach-Zehnder modulator. AWG: Arbitrary waveform generator. PC: Polarization controller. LF: Lensed fiber. EDFA: Erbium-doped fiber amplifier. BPF: Bandpass filter. PR: Photoreceiver. OSA: Optical spectrum analyzer.}
\label{fig_2}
\end{figure}

It is important to note that choosing the right detuning is key to our photonic filter approach. In particular, our filtering scheme relies on partial injection locking of the MWL to the optical carrier at $\lambda_\text{c}$ \cite{abdollahi_mode-coupling_2022}. In partial locking, the gain of the MWL is not completely depleted by the injection at $\lambda_\text{c}$, but the gain is shared with the other modes of the MWL. In addition, the signals at $\lambda_\text{I}$ and $\lambda_{\text{II}}$ will experience regenerative amplification whose selectivity is controlled via the feedback phase, enabling signal filtering. Partial locking is achieved by injecting the carrier at the low detuning values of the locking range. In Section V, we describe and investigate in detail the impact of the detuning. 

The emission from the MWL is extracted through an optical fiber circulator (OFC). We measure the optical spectrum with an optical spectrum analyzer (OSA), and the time domain signal is captured with an oscilloscope (Teledyne Lecroy WavePro HD) connected to a photoreceiver (PR, Thorlabs RXM42AF) after filtering out the optical carrier with a bandpass filter (BPF) with approximately 40 GHz 3-dB bandwidth. Note that in case of considering the O-T conversion, no BPF would be used and the carrier would serve as the local oscillator to generate the THz signal on a, e.g., high-speed photodector \cite{nellen_experimental_2020}. An erbium-doped fiber amplifier (EDFA) set to an output power of 5~dBm is employed to compensate for the coupling losses and the insertion loss from the BPF. The signals from the oscilloscope are processed to extract the bit error ratio (BER), obtaining values below 10-10 for the lowest $P_{\text{sig}}$ of ‒ 46~dBm, see inset in Fig.~2(a) for the corresponding eye diagram. Fig.~2(c) depicts the input (top) and output (bottom) optical spectrum of the signal when the photonic filter is set to select the channel at $\lambda_\text{I}$. We measure a suppression ratio between the power $P_\text{I}$ of the signal at $\lambda_\text{I}$ and the power $P_{\text{II}}$ of the signal at $\lambda_{\text{II}}$ ($P_{\text{I}}/P_{\text{II}}$) of up to 12.3~dB and an optical gain of up to 13~dB. Both power levels and optical gain are on-chip, estimated by considering an LF to chip coupling losses of 3~dB. Figures~2(d) and (e) show the dependency of the suppression ratio and optical gain as a function of the optical power of the injected signal. Due to the gain saturation in our MWL, injecting low signal powers leads to higher gains.

\IEEEpubidadjcol

\section{FEEDBACK-CONTROLLED PHOTONIC FILTER}

We demonstrate agile control of the photonic filter and characterize the filtering in terms of the power suppression ratio, optical gain, and 3-dB bandwidth. To this end, we generate a broad signal, 1.3~THz offset from the optical carrier, which we emulate using amplified stimulated emission (ASE) noise. Fig. 3(a) shows the employed setup: The broad signal is emulated with an ASE noise source and a BPF with approximately 100~GHz of 3~dB bandwidth, see Fig.~3(b). Similar to the experimental setup described in Fig.~2(a) of Section~III, we combine the optical carrier and the signal and inject them into the MWL. The spectral components of the signal at $\lambda_\text{I}$ and $\lambda_{\text{II}}$, which coincide with modes $\lambda_1$ and $\lambda_2$ of our MWL, respectively, undergo regenerative amplification. By changing the voltage $V_{\text{PM}}$ applied to the PM inside the feedback cavity, see Fig.~1, we are able to control the optical powers $P_{\text{I}}$ and $P_{\text{II}}$ of each component. Figure~3(c) depicts the suppression ratio $P_{\text{I}}/P_{\text{II}}$ (top) and optical gain (bottom) as a function of the voltage $V_{\text{PM}}$. For the signal component at $\lambda_\text{I}$, we measure, at $V_{\text{PM}} = ‒ 4.5$~V, a suppression ratio of 19.8~dB, an optical gain of 16.3~dB, and a 3-dB bandwidth of approximately 160~MHz. For the component at $\lambda_{\text{II}}$, we measure, at $V_{\text{PM}} = ‒ 8.25$~V, a suppression ratio of -14 dB, an optical gain of 12.5~dB, and a 3-dB bandwidth of approximately 200~MHz. In this configuration, for $V_{\text{PM}}$ values between approximately ‒ 3.25~V and ‒ 6.25~V, our photonic filter selects the signal at $\lambda_\text{I}$, while between ‒ 7.75~V and ‒ 9.25~V, we are able to adjust the filter to select the signal at $\lambda_{\text{II}}$, with a suppression ratio of more than 10~dB. Figure 3(c) shows exemplary output spectra for $V_{\text{PM}}=-4.5$~V (top) and $V_{\text{PM}}=-8.25$~V (bottom). 

Note that, although in this work we adjust the voltage values stepwise, the phase modulator allows fast switching between the two MWL emissions \cite{xu_advanced_2019}. In addition, with our feedback-controlled approach and setting $V_{\text{PM}} = ‒ 7$~V, we are able to amplify both components equally with a gain of 9~dB.

\begin{figure}[!t]
\centering
\includegraphics[width=3.6in]{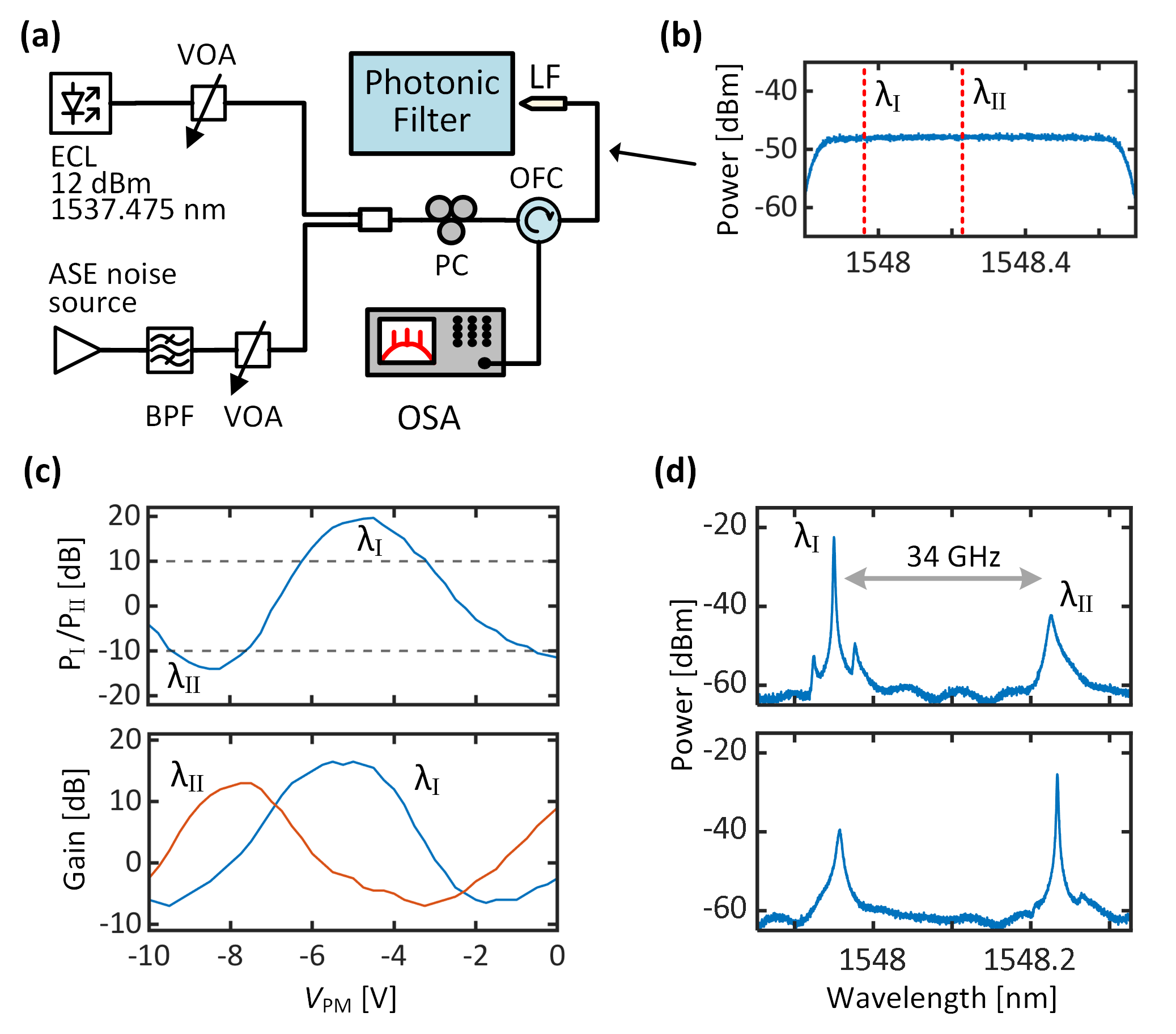}
\caption{Feedback-controlled photonic filter. (a) Setup for the generation and filtering of a broadband optical signal. The 100~GHz wide optical signal is generated with an amplified stimulated emission (ASE) noise source, followed by a bandpass filter (BPF). The signal is combined with an optical carrier from an external cavity laser (ECL) offset by 1.3~THz to emulate the THz-to-optical conversion of a 100 GHz wide THz signal. A polarization controller (PC) is used to align the polarization of the ECL with the MWL. (b) Power spectrum of the ASE noise measured before injection in the MWL. (c) Top: suppression ratio or ratio between the power of the channel at $\lambda_\text{I}$ and that of the channel at $\lambda_{\text{II}}$. Bottom: Optical gain associated with each of the channels. (d) Optical spectra for $V_{\text{PM}}= -4.5$~V (Top) and $V_{\text{PM}} = -8.25$~V (Bottom).}
\label{fig_3}
\end{figure}

\section{STABILITY ANALYSIS OF THE PHOTONIC FILTER}

The properties of our photonic filter depend on various laser and signal parameters. In this section, we first analyze the impact of the power of the injected signal $P_{\text{sig}}$ and the detuning $\delta$ for a constant carrier power $P_c$, and then analyze the impact of the carrier power.

\subsection{Gain saturation }

The gain of our photonic filter directly depends on the injected signal power, $P_{\text{sig}}$. Low input powers yield high optical gain—up to nearly 20~dB for the weakest injected signals, see Fig.~4(a), where power values are on-chip. This is attributed to gain saturation in the MWL: as the channel’s power approaches the saturation power, the available gain for further amplification decreases. Consequently, higher input power values not only feature lower gain but also lower suppression ratio values, as the feedback-controlled mode competition becomes less effective at suppressing the non-selected channel. Gain saturation also leads to an increase in the 3~dB bandwidth of the photonic filter, as shown in Fig.~4(a), bottom panel. Note that the gain is already depleted by the carrier, therefore, we expect the power of the carrier to impact the achievable gain and suppression ratio, see the next section.

Regarding THz communication links, optical signals after the T–O conversion are typically weak  \cite{ummethala_thz--optical_2019}. Therefore, the regenerative amplification provided by our filter can significantly enhance the signal-to-noise ratio prior to optical-to-THz conversion. While in this work we fix the current applied to the SOA in the feedback cavity to 25~mA, increasing this current—and thus the feedback strength—could potentially raise the maximum achievable suppression ratio for higher signal powers, which is a promising direction for future optimization.

\begin{figure}[!t]
\centering
\includegraphics[width=3.6in]{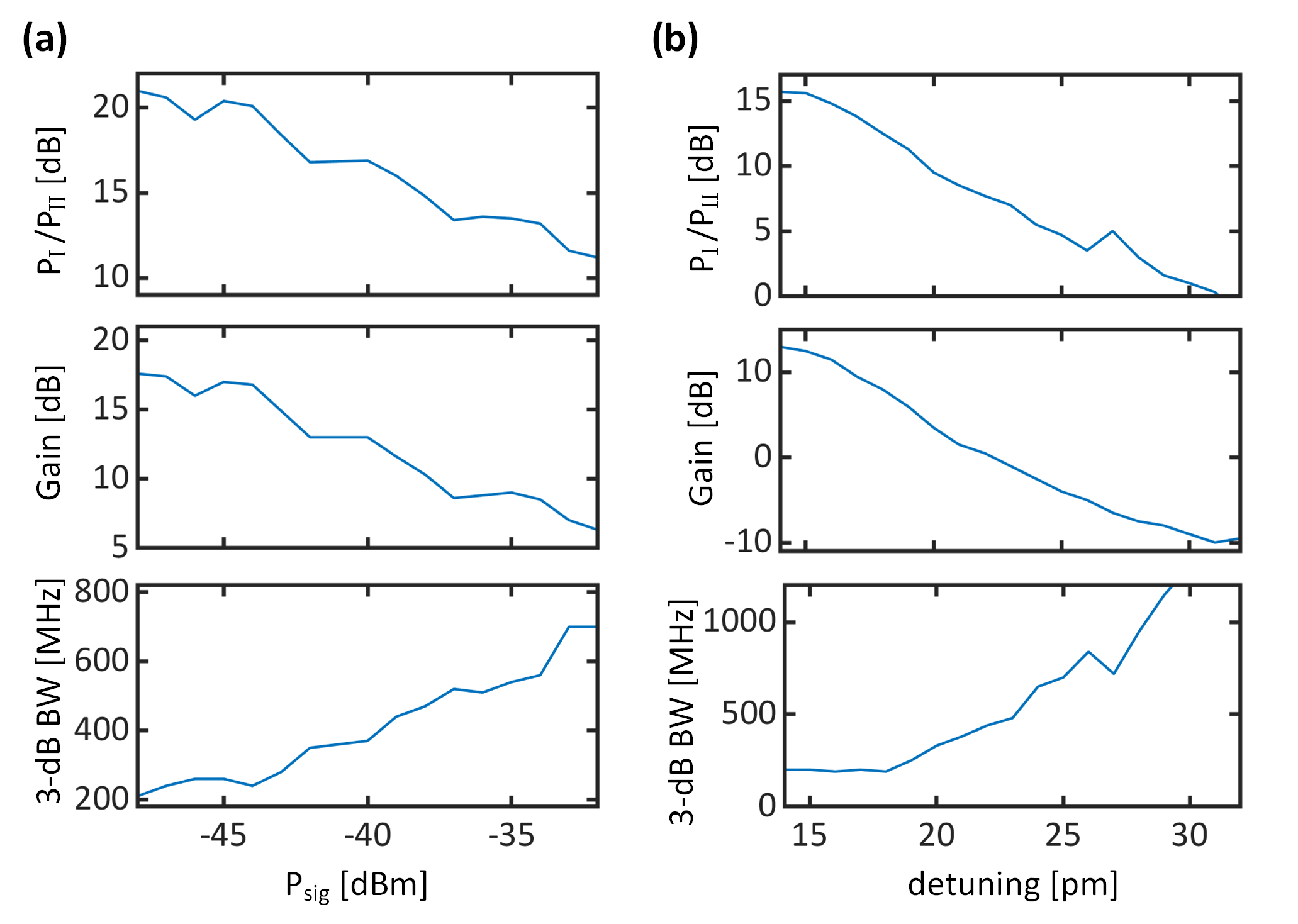}
\caption{Characterization of our photonic filter for $V_{\text{PM}} = -4$~V and $P_{c} = 3$~dBm. (a) Suppression ratio, defined as the ratio between the power of the channel at $\lambda_\text{I}$ and that of the channel at $\lambda_{\text{II}}$ (Top), gain (Middle), and 3-dB optical bandwidth (Bottom) as a function of the on-chip power of the optical signal measured before the lensed fiber. Detuning $\delta = 15$~pm. (b) Suppression ratio (Top), gain (Middle), and 3-dB optical bandwidth (Bottom) as a function of the relative detuning of the optical carrier’s wavelength with respect to that of the MWL. $P_{\text{sig}} = -40$~dBm.}
\label{fig_4}
\end{figure}

\subsection{Carrier power and detuning}

Our filtering scheme relies on partial injection locking of the MWL to the optical carrier at $\lambda_\text{c}$ \cite{abdollahi_mode-coupling_2022}, as described in Section III. Stable locking is achieved by finely adjusting the wavelength detuning $\delta = \lambda_\text{c} - \lambda_0$ to operate within the partial locking region of the injection-locking range. Figure~4(b) shows the effect of detuning on suppression ratio, optical gain, and 3~dB bandwidth. From these plots, we observe that there is an optimum range of detuning values. In Fig.~5, we explore this range by mapping the suppression ratio and the optical gain as a function of the detuning and carrier power. The colored values indicate the region of locking, while the white area is associated with unlocked states, such as dynamics, typically at low detuning values \cite{abdollahi_mode-coupling_2022}.

For low detuning values, the MWL is close to the edge of the locking range, where partial locking appears, consequently, the signals $\lambda_\text{I}$ and $\lambda_{\text{II}}$ have access to a larger share of the pool of carriers. In combination with the selectivity introduced by the feedback cavity, it enables a larger suppression ratio and optical gain. However, excessive detuning risks falling outside the locking range, leading to dynamic instabilities and degraded signal quality. On the other side, large detuning values within the locking range may lead to complete locking, i.e., the optical carrier depletes most of the gain of the MWL and little or no power is observed at $\lambda_\text{I}$ and $\lambda_{\text{II}}$.  

\begin{figure}[!t]
\centering
\includegraphics[width=3.5in]{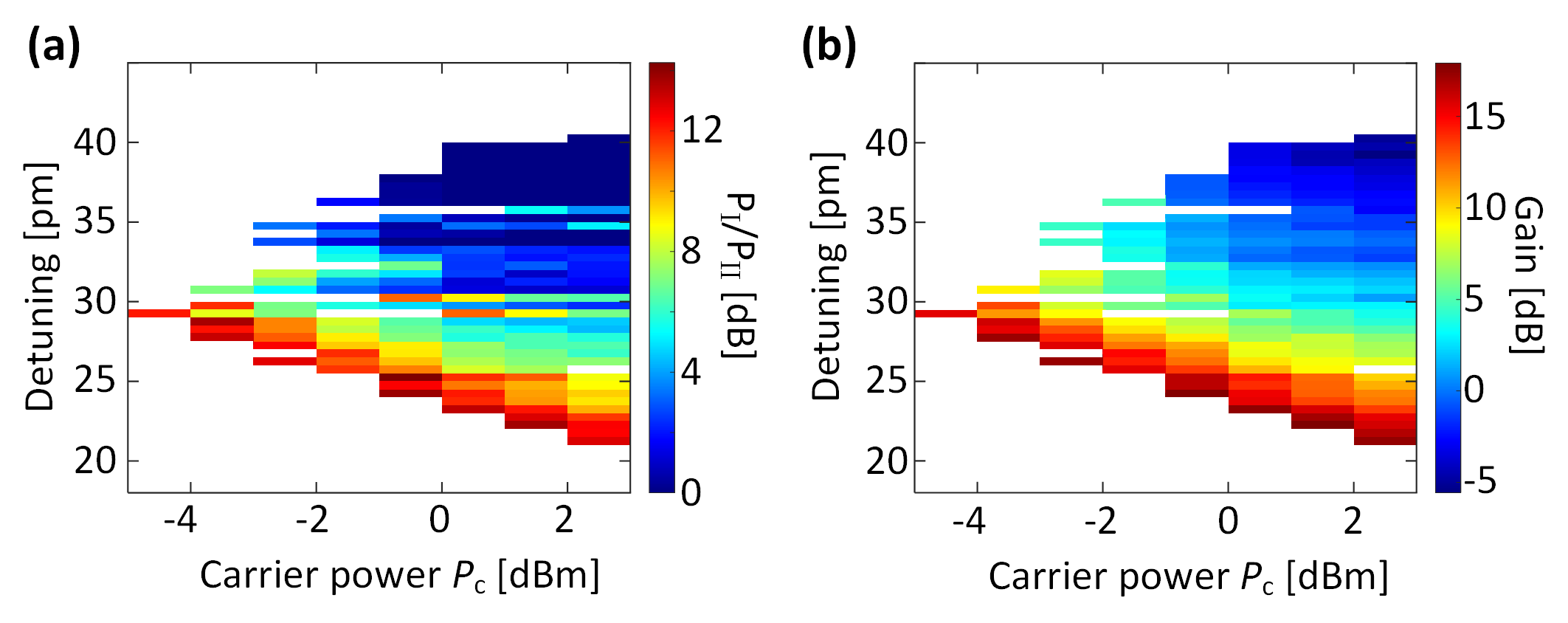}
\caption{Stability analysis of the photonic filter for $f_{\text{THz}} = 1.3$~THz. (a) Power suppression ratio and (b) optical gain as a function of the detuning and the carrier power for $V_{\text{PM}}= -3.3$~V, $P_{\text{sig}} = - 40$~dBm.}
\label{fig_6}
\end{figure}

Note that due to the non-zero linewidth enhancement factor of the semiconductor MWL, optical injection will red shift the resonance frequency $\lambda_0$0 in such a way that, in general, positive detuning values are required for locking the MWL and the optimum detuning value, i.e., leading to the highest suppression ratio, will change with the optical power of the injected carrier as it is shown in Fig. 5. In Figs.~5(a) and (b) we keep $P_{\text{sig}}= – 40$~dBm, and $V_{\text{PM}}=-3.3$~V. We did not optimize $V_{\text{PM}}$ for every detuning and carrier power; doing so could further improve the filter performance. Importantly, the carrier is also present at the output of the MWL, which is required for optical-to-THz conversion.

\section{PHOTONIC FILTER FOR 270 GHZ SIGNALS}
In this section, we make use of a second MWL in which the pitch of the DBR components is set so that $f_{\text{THz}} = 270$~GHz. In this case, the modes of the MWL emit at $\lambda_0 = 1546.55$~nm, $\lambda_1 = 1544.29$~nm, and $\lambda_2 = 1544.52$~nm, with a channel spacing of approximately 29~GHz. We drive the MWL with 40~mA and set it to 22~°C, the current applied to the SOA in the feedback cavity is 25~mA. We use a similar setup to that of Fig.~3(a), and we also obtain suppression ratio values of more than 10~dB for both signals and gain values of up to 10~dB, see Fig.~6.

\begin{figure}[ht]
\centering
\includegraphics[width=1.9in]{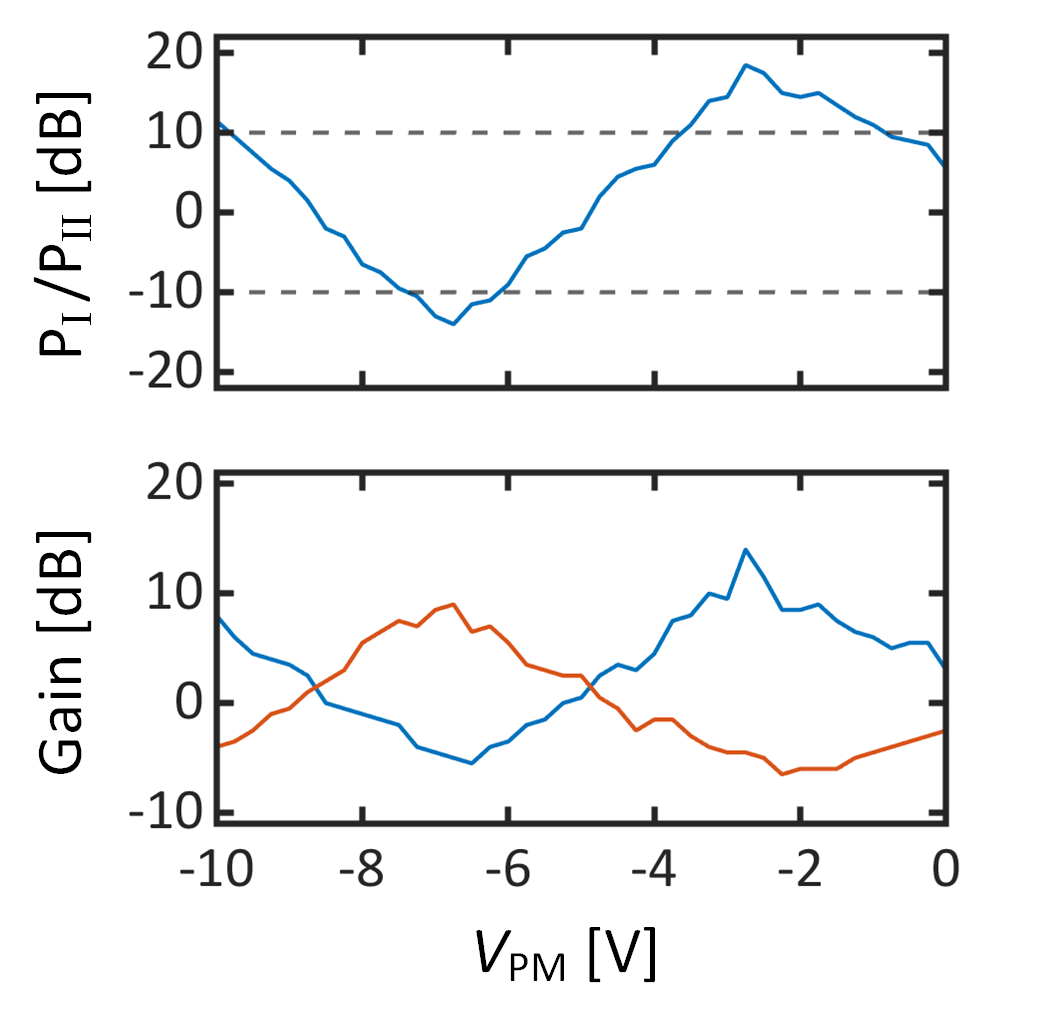}
\caption{Photonic filter for 270 GHz signals. Suppression ratio (top) and gain (bottom) as a function of the voltage $V_{\text{PM}}$ applied to the phase modulator of the feedback cavity of an MWL with $f_{\text{THz}} = 270$~GHz.}
\label{fig_5}
\end{figure}

\section{Conclusion}

We have demonstrated a narrowband, tunable photonic filter for THz communication links based on optical signal injection into a feedback-controlled integrated multi-wavelength laser. By adjusting the feedback phase using an on-chip electro-optic phase modulator, the filter can selectively amplify or suppress specific spectral channels with nanosecond switching capability. The device achieves suppression ratios up to 20 dB, optical gains of more than 15 dB for low-power signals, potentially removing the need for additional components such as external amplifiers, avoiding the associated amplified spontaneous emission (ASE) noise. In addition, it features 3 dB bandwidths as narrow as 160 MHz. While the level of suppression ratio is still below that provided by CROW filters \cite{taddei_high-selectivity_2019} or stimulated Brillouin-based filters \cite{liu_integrated_2020}, our filter provides bandwidth values comparable to those provided by other types of photonic filters without complex tuning mechanism. However, unlike conventional passive filters, our approach provides regenerative amplification, making it particularly well-suited for weak optical signals after THz-to-optical conversion, and fast reconfigurability. The demonstrated agility, gain, and narrowband operation highlight the potential of this architecture for high-capacity, reconfigurable THz wireless communication systems.

We have also explored our approach for signals at 270 GHz, achieving similar values of suppression ratio, gain, and bandwidth, demonstrating the feasibility of our scheme for lower THz signals.

The proposed architecture also holds promise as a demultiplexer for satellite communications, where filtering of signals with bandwidths of hundreds of MHz is required \cite{mitsolidou_microwave_2023}. While satellites typically operate at lower RF frequencies around 30–50 GHz, our 270 GHz demonstration confirms that the resonance frequencies of the MWL can be flexibly designed. This capability could enable compact, reconfigurable DEMUX filters with high selectivity for next-generation satellite payloads.

\section*{Acknowledgments}
The authors gratefully acknowledge professor Martin Virte for his invaluable guidance and insightful comments on the manuscript.

\bibliography{2025_Photonic_filter_BibTeX}
\bibliographystyle{IEEEtran}

\end{document}